\documentclass[a4paper,11pt]{article}
\pdfoutput=1 

\usepackage{jcappub} 

\usepackage[T1]{fontenc} 
\usepackage{dirtytalk} 

\title{The Exochronous Universe: \\a static solution to the Einstein field equations}

\author{Robin Booth}

\affiliation{Astronomy Centre, School of Mathematical and Physical Sciences, \\University of Sussex, Brighton BN1 9QH, United Kingdom}

\emailAdd{robin.booth@sussex.ac.uk}

\abstract{
Arguably our current cosmological paradigm, the so-called $\Lambda$CDM `concordance model', faces an existential crisis. This has largely been brought about by its reliance on the twin concepts of dark matter and dark energy, and the continued inability of the observational and theoretical physics community to find viable candidates for these postulated phenomena.  While it is still possible that this search will eventually prove successful, it is perhaps worthwhile looking at alternatives, and in particular, re-examining the very foundations of our current cosmological model to see whether an entirely new cosmological paradigm might provide a better explanation.

The main failures of the prevailing cosmological paradigm were reviewed, concentrating on the coincidences and contradictions presented by the relationship between dark matter and dark energy.  Revisiting the core features of this paradigm,  the Friedman-Lema\^itre-Robertson-Walker (FLRW) metric is identified as the likely root cause of current issues with the $\Lambda$CDM model.  Building on the concept of a `timeless' universe suggested by a number of theorists in recent years, the exochronous metric is introduced and is shown to give rise to a static solution to the Einstein field equations. The cosmological implications of this model are explored, and we demonstrate that it can result in a universe with a critical energy density that is close to the currently observed baryonic matter density, without the need to invoke dark matter or dark energy.  
}

\begin{document}
\maketitle
\flushbottom

\section{Introduction}

The so-called  $\Lambda$CDM `Concordance Model' of cosmology is currently our best attempt to describe the origin, evolution, and dynamics of the universe. (See \cite{Springel2006} for an overview).  There is considerable prima facie evidence to support the main constituents of this model, namely `dark energy' in the form of a cosmological constant($\Lambda$), and cold non-baryonic `dark matter'. However, this model is not without problems. Indeed, the continuing inability of the scientific community to identify the origins of dark energy and dark matter arguably constitutes the two biggest unresolved questions in physics today.

In the remainder of this section we briefly review the evidence for the various elements that constitute the concordance model, including the cosmological constant and dark matter. Section \ref{sec:problems} outlines some of the main issues inherent in the $\Lambda$CDM model, including the apparent cosmological coincidences that arise from the model's parameters.  Section \ref{sec:history} provides a historical overview of the steps that led to Einstein's description of his static universe, and the subsequent formulation of the FLRW metric and the expanding universe that it describes.  In Section \ref{sec:time_in_GR} we outline some of the competing proposals for dealing with time within the GR framework.   This is followed in Section \ref{sec:exochronous} by proposing a new timeless metric that results in a static solution to Einstein field equations. Finally, in Section \ref{sec:conclusions} we discuss the implications of this timeless metric in terms of its ability to resolve some of the issues identified in section \ref{sec:problems}.

\subsection{Hot Big Bang cosmology}
The current hot big bang (HBB) cosmological paradigm, in which the concordance model forms an integral part, contains several ingredients that are summarised here.

\subsubsection{General Relativity}
An underlying assumption in the following discussion is that general relativity (GR) is the correct theory of gravitation to be employed in considering the structure and dynamics of the universe as a whole, and of its constituent components.  In this context we define GR in terms of the Einstein gravitational field equations.

\begin{equation}\label{eqn:efe}
    G_{\mu\nu} + \Lambda g_{\mu\nu} = \frac{8 \pi G}{c^4} T_{\mu\nu} 
\end{equation}

where
\[ G_{\mu\nu} = R_{\mu\nu}  - \tfrac{1}{2}R g_{\mu\nu} \]

defines the Einstein curvature tensor in terms of the Ricci tensor and Ricci scalar.

\subsubsection{FLRW Metric}
A number of viable metrics can be used to construct the Ricci tensor from which to generate solutions to Equation \ref{eqn:efe}, of which the Schwarzchild metric describing the curvature of spacetime in the vicinity of an isolated spherically symmetric body is possibly the best tested, for example, in solar system tests.  However, the prevailing metric used to solve Equation \ref{eqn:efe} when considering the bulk behaviour of the universe as a whole is the Friedman-Lema\^itre-Robertson-Walker (FLRW) metric.  The history of the genesis of this metric from Einstein's original static universe is reviewed in more detail in section \ref{sec:history}.

\subsubsection{Energy-Momentum}
If one treats the cosmological constant as a vacuum energy, that is, lying on the right-hand side of Equation \ref{eqn:efe}, then the three main components contributing to the Friedman equations are:
\begin{itemize}
    \item conventional baryonic matter
    \item cold dark matter
    \item vacuum energy (dark energy)
\end{itemize}
The evidence for each of these components of the the $\Lambda$CDM concordance model is briefly reviewed in section \ref{sec:concordance}.

\subsubsection{Inflation}
Although inflation is not a component of the Friedman equations, and is as yet only a hypothesis, it is nevertheless a significant part of the prevailing cosmological paradigm as it is capable of solving several problems associated with the Big Bang, including the flatness and horizon problems (see \cite{Guth1981} for details).

\subsection{Ingredients of the \texorpdfstring{$\Lambda$CDM}{LambdaCDM} Concordance Model \label{sec:concordance}}

\subsubsection{Baryonic Matter}
Our best estimate from astronomical observations and analysis of the CMB is that baryonic matter constitutes no more than 4.5\% of the critical energy density of the universe, of which only approximately 0.5\% is made up of luminous baryonic matter contained in stars and galaxies.

\subsubsection{Dark Matter \label{sec:CDM}}
The history of dark matter dates back to 1933, when its existence was inferred by Zwicky from the dynamics of galaxy clusters (see \cite{VandenBergh1999} for example).  Since then, evidence for the pervasive presence of dark matter has become overwhelming and includes galactic rotation curves, the structure of galaxy groups and clusters, large-scale cosmic flows, and gravitational lensing.  

\subsubsection{Dark Energy}
The possibility that matter (both baryonic and dark) might only constitute a relatively small proportion of the total energy density of the universe predates the 1998 discovery of the acceleration in the cosmic expansion rate. The existence of a nonzero cosmological constant was first postulated in 1990, based on power spectrum measurements of large-scale structures obtained from galaxy surveys \cite{Efstathiou1990}.

The expansion history of the universe can be explored by measuring the relationship between the luminosity distance and the redshift for a light source with a known intrinsic magnitude.
Just such an ideal `standard candle' has been identified in the form
of Type Ia supernovae (SNe Ia). Several studies have been conducted by
various groups, including the Supernova Cosmology Project \cite{Perlmutter1999}
and the High-Z Supernova Search Team \cite{Riess1998}, to
measure the distances of a relatively large sample of supernovae at
redshifts extending up to $z>2$.  In 1998 these research teams independently identified an apparent acceleration in the cosmic expansion rate, commencing at an epoch corresponding to a redshift of $z \simeq 0.5$.  

The generally favoured candidate for the origin of this acceleration is a cosmological constant, corresponding to the $\Lambda$ term in the Einstein field equations of general relativity.  In the context of the $\Lambda$CDM model, calculations indicate that the best fit with the observed SNe Ia results is obtained with $\Omega_\Lambda \simeq 0.7$ and $\Omega_m \simeq 0.3$, where $\Omega_\Lambda, \Omega_m$ are the respective contributions of the cosmological constant and all forms of matter to the critical energy density $\Omega$.  

\subsection{Supporting Evidence}

\subsubsection{Cosmic Microwave Background}
In addition to the evidence for dark matter and dark energy described above, other sources would appear to corroborate the existence of these two phenomena.  The most significant of these are the measurements of the cosmic microwave background (CMB) by the successive COBE, WMAP and Planck satellite missions.  The results from the analysis of the most recent Planck dataset \cite{Planck2015} gave a value for $\Omega_m = 0.308$, with a spatial curvature of zero, implying that $\Omega_{tot}=1$ and that $\Omega_\Lambda = 0.69$.  These are in close agreement with the values obtained from the SNe Ia measurements, hence justifying the description of the current $\Lambda$CDM cosmology as the `Concordance Model'.

\subsubsection{Baryon Acoustic Oscillations}
Another useful tool capable of acting as a `standard ruler' is the use of baryonic acoustic oscillations (BAO) measurements.  These link density fluctuations that occurred at the epoch of the early universe sound horizon with large-scale structure post-CMB. \cite{Basset2010}.  They provide another cosmological standard ruler capable of linking redshift with distance and, as such, provide an independent set of observational data that is capable of supporting SNe Ia measurements.

\section{Problems with \texorpdfstring{$\Lambda$CDM}{LambdaCDM} Cosmology \label{sec:problems}}

Despite its many successes, the $\Lambda$CDM model suffers from a number of issues, which individually may not be considered fatal, but which collectively add up to a substantial body of circumstantial evidence against this being the correct cosmological model.  These issues can be divided into four broad categories:
\begin{itemize}
    \item Fine tuning problems
    \item Ontological issues
    \item Metaphysical considerations
    \item Conflicting observational datasets
\end{itemize}
These flaws are now discussed in more detail in the remainder of this section.

\subsection{Fine Tuning Problems\label{sec:fine_tuning}}
The concordance model provides estimates for the constituent components of the critical energy density of $\Omega_{bm} = 0.049$, $\Omega_{dm} = 0.268$, and $\Omega_\Lambda = 0.683$ (from the Planck Consortium 2015 results \cite{Planck2015}), where
$\Omega_{bm}$ and $\Omega_{dm}$ are the proportions of baryonic matter and dark matter, respectively, 
\[\Omega_{tot} = \Omega_{bm} + \Omega_{dm} + \Omega_\Lambda = 1\]
\[	\Omega_{M}= \Omega_{bm} + \Omega_{dm}\equiv\left({\frac{{8\pi G}
}{{3H_{0}^{2}}}}\right)\rho_{0}\]
\[	\Omega_{\Lambda}\equiv\frac{\Lambda}{{3H_{0}^{2}}}\]

\subsubsection{The Flatness Problem}
The first fine-tuning problem is known as the flatness problem: Why is the present day density of the universe $\rho_0$ so close to the critical value required for $\Omega=1$?  Cosmic inflation has been postulated as a solution to this problem \cite{Guth1981}; however this mechanism fails adequately to explain how it acts on the three \textit{individual} components that constitute $\Omega$ to make $\Omega=1$.

\subsubsection{The Linearity Problem}
The second fine-tuning problem is less well documented \cite{Kutschera2006}, but is equally intriguing: why is the age of the universe $\approx 1/H_0$, where $H_0$ is the present-day value of the Hubble parameter?  To restate the issue more succinctly, why are the observed proportions of $\Omega_{bm}$, $\Omega_{dm}$ and $\Omega_\Lambda$ \textit{precisely} those that give rise to a linearly expanding universe, similar to the so-called empty Milne universe in which the age does indeed $=1/H_0$ ?  Another way to state this problem is to note the observation that the present-day values of  $\Omega_{m}$ and $\Omega_{\Lambda}$ are both of order unity whereas in the past and in the future they will differ by many orders of magnitude due to the  $\Omega_{m}a ^3$ scaling  as the universe expands.   If we exclude anthropic arguments, then the chances of this arising by coincidence must be extremely small, which suggests the existence of some currently unexplained underlying mechanisms. 

\subsection{Ontological Considerations \label{sec:what}}
$\Lambda$CDM cosmology depends on the existence of two key components to supply the parameters for the Friedman equations that govern the dynamics of the universe: dark matter and dark energy.  Additionally, a third ingredient, the inflaton,  is required to generate the inflationary potential field that is postulated to drive the initial exponential expansion of the cosmos, and to give rise to the primordial matter power spectrum that will ultimately determine the evolution of large-scale structure in later cosmological epochs.  

\subsubsection{Dark Matter}
Various standard model candidates for dark matter have been proposed; in particular, the existence of various additional species of neutrinos. Other possible particulate candidates include WIMPs, axions, MACHOs, and particles emerging from supersymmetry theories, such as gravitinos.  A comprehensive summary of the dark matter universe that describes these and many other candidates is provided in \cite{Bertone2005}. However, despite of the attempts of numerous research projects to detect these hypothetical particles, to date none has proven successful, and the origins of dark matter remain as elusive as ever.  Another possible dark matter candidate that has been examined more recently is primordial black holes \cite{Carr2016}.  While it is not yet possible to rule out this source, the probability of it being the entire answer to the dark matter problem is considered minimal.

\subsubsection{Dark Energy}
The most obvious explanation for the cosmological constant is thought to be the vacuum energy arising from quantum loop corrections at the Planck scale, which is of the order of $10^{19}$ GeV, where gravity should unify with the other fundamental forces. However, the value of $\Lambda$ required to account for the observed cosmological acceleration is a factor of $10^{123}$ smaller than this, which is a non-insignificant discrepancy in need of an explanation! Numerous alternative gravitation theories have been proposed as replacements or enhancements to GR, some of which are claimed to be able to reproduce the effects of a cosmological constant, for example $f(R)$ gravity models.  A useful review of some of the possible forms of $\Lambda$ is provided in \cite{Durrer2008}.

\subsubsection{The Inflaton}
Although the inflaton and its associated field are not in themselves components of the $\Lambda$CDM model, they are mentioned here for the sake of completeness in reviewing the entire set of ingredients that constitute the current cosmological paradigm.  The main criticism that can be levelled at the inflationary paradigm is that there are no obvious candidates for the inflaton field, and as a consequence there are almost as many variants of inflation potential as there are researchers working in this field.  

\subsection{Metaphysical Considerations}

\subsubsection{The metric is not the universe}
One philosophical issue with the FLRW metric is that it conflates the concept of a metric, essentially a framework for measuring the separation between points in metric space,  with the notion of a physical universe.  In the way it is used as one of the components of the gravitational field equations, particularly with respect to the cosmological constant, there is the implication that there exists some physical mechanism that causes the metric, and therefore the universe, to expand.

\subsubsection{The metric should be viable in the absence of clocks and rulers}
Simply stated, any generalised metric that can be applied to the universe as a whole (as opposed to a metric that is restricted to a well-defined mass distribution in a confined volume of space, such as the Schwarzchild metric), must be able to do its job of specifying the separation of points in space in the limit where the energy density of the universe it describes tends to zero.  At this point the universe will contain no particles or fields that can act as clocks and rulers to measure the scale.  One case in point is the so-called Milne empty universe  \cite{Milne1933}, in which the radius of the universe is found to expand at the speed of light, $r=ct$. The only way to measure the separation of points in space, and the curvature of that space, is in terms of angular separation. Or to put it another way, protractors are the only measuring device that will work as effectively in an empty universe as in one that contains matter or radiation.  We can summarise this argument with the statement that, logically, the metric cannot possess an absolute scale.

\subsection{Hubble constant tension \label{sec:H_tension}}
Until recently it could be said that all observational data supported the $\Lambda$CDM model with a 1\% level of accuracy.  However, recent high precision measurements of the present day value of the Hubble parameter, $H_0$, on local distance scales, using a combination of Cepheids, SNe Ia and BAO data, have identified a $ \sigma > 4.5$ tension between this value ($ \simeq \, 73 \, km \, s^{-1} \, Mpc^{-1}$) and the  model-dependent value inferred from the Cosmic Microwave Background ($ \simeq \, 67 \, km \, s^{-1} \, Mpc^{-1}$). \cite{Bernal2016}.  This degree of tension between the two determinations of $H_0$ can  only be realistically resolved using (a) changes to the parameters determining the early stage ($z>1000$) evolution of the universe, and/or (b) modifications to early universe physics that would impact on the sound horizon at recombination, or (c) a more significant revision of the underlying $\Lambda$CDM cosmological model.

\section{Historical Background to \texorpdfstring{$\Lambda$CDM}{LambdaCDM} Cosmology \label{sec:history}}

\subsection{The Einstein Static Universe \label{sec:Einstein_static}}
Arguably, Einstein's 1917 paper, `\textit{Cosmological Considerations in the
General Theory of Relativity}' \cite{Einstein1917} marks the birth of modern cosmology, providing a relativistic treatment of the dynamics of the universe.
A comprehensive review of this paper is provided in \cite{ORaifeartaigh2017} so we restrict our discussion here to a summary of the essential steps that led to his conclusions. (Another informative perspective on the development of cosmology in its early years was provided by Milne in \cite{Milne1933}).

To understand the true extent of Einstein's cosmological insights displayed in this paper it is necessary to begin with the commonly held  picture of the universe that prevailed at the time of writing.  Essentially, this model assumed that the Milky Way galaxy comprised almost the entirety of our universe, with the remainder consisting of sparsely distributed extra-galactic nebulae of unidentified composition. Furthermore, it was assumed that the universe in which our galaxy was embedded was infinite, with the matter density dropping to zero at large distances from the galactic centre. The problem Einstein was presented with in trying to apply general relativity to this cosmological model was in trying to construct a metric that adequately handled the boundary conditions, with the gravitational potential decaying to zero at infinity.  His solution to this problem was to postulate a spatially closed universe with a uniform matter density, which was homogeneous and isotropic on the largest scale. This description, adhering to what we now term the Cosmological Principle, corresponds closely to the present-day model of the cosmos, and is now supported by over a century of astronomical observations since Einstein made this conceptual leap of the imagination.

In considering the metric that should be used to construct the Einstein curvature tensor $G_{\mu \nu}$ used on the left-hand side of the gravitational field equations, Einstein naturally started with the concept of a hyper-surface of constant curvature, defined by \footnote{in Einstein's day the convention was to use $x_4$ as the time dimension, as opposed to the present day convention of using $x_0$}:
\[
R^2 = \xi^2_1 + \xi^2_2 + \xi^2_3 + \xi^2_4
\]

Again, perhaps understandably in view of his history of using the Minkowski metric in the Special Theory of Relativity, Einstein  chose to use time as one of the four dimensions in the metric even though this is not a prerequisite for general relativity, as discussed below in section \ref{sec:time_in_GR}.  The line element in the Einstein metric, expressed in spherical coordinates, is given by

\[ ds^{2}=-c^{2}\mathrm{d}t^{2}+\mathrm{d}r^{2}+r^{2}\mathrm{d}\theta^{2} + r^2 \sin^2 \mathrm{d}\phi^2 \]

which in elliptical coordinates becomes

\begin{equation} \label{eqn:Einsteinstatic}
ds^{2}=-c^2dt^{2}-R^{2}\left({\frac{{dr^{2}}}{{1-kr^{2}}}+r^{2}d\theta^{2}+r^{2}\sin^{2}\theta d\phi^{2}}\right)
\end{equation}

(It can be noted that this 4D metric reduces to the Minkowski metric in flat space when $k=0$.)

When Einstein attempted to use this metric to construct the Ricci tensor on the right hand side of the gravitational field equations, he found that this resulted in infinities for the spatial components of the tensor equation when the pressure terms approached zero. (See \ref{app:Einstein_metric} for a detailed derivation of this non-solution).  His proposed solution was to insert a cosmological constant term, $-\lambda g_{\mu \nu}$ into the curvature side of the field equation: \footnote{This is the form of the equation used in Einstein's original paper. $T$ here is the trace of the stress-energy tensor, the inclusion of which is not explicitly explained in the original text.  However, its presence makes no difference to the general outcome of the calculation. }

\begin{equation}
    G_{\mu\nu} -\lambda g_{\mu \nu}  = -\kappa \left(T_{\mu\nu}  - \tfrac{1}{2}T g_{\mu\nu} \right)
\end{equation}

where  $\kappa = \frac{8 \pi G}{ c^2}$  is the so-called Einstein constant.
 With the inclusion  of this cosmological constant, it is then possible to solve the gravitational field equations (see \ref{app:Einstein_metric}), giving
 
 \begin{equation}
    R = \frac{1}{\sqrt{\lambda}}
\end{equation}

where $R$ can be thought of as the radius of the cosmos, and

\begin{equation}
   \lambda = \frac{4 \pi G \rho}{c^2}
\end{equation}

From Einstein's point of view, this was a satisfying result because it showed that it was indeed possible to apply his general theory of relativity to a cosmological model that was consistent with the prevailing picture of a static universe.  The fact that this model was unstable with respect to small perturbations in the matter density field was not noticed at that time.

\subsubsection{de Sitter Space}

For the sake of completeness it is perhaps useful to briefly mention de Sitter's response to \cite{Einstein1917}, contained in \cite{DeSitter1917}. In this study, de Sitter identifies an alternative solution to the modified gravitational field equations (with the included $\lambda)$ in the case of a universe that is devoid of matter, that is, $\rho=0$.  This results in the following solution
 \begin{equation}
    R = \frac{3}{\sqrt{\lambda}}
\end{equation}

\subsection{The Friedman-Lem\^{a}itre-Robertson-Walker Metric \label{sec:FLRW}}

The next major development in applying GR to cosmology comes from Friedman's 1922 paper \cite{Friedman1999}, in which the author, in addition to considering the generalised solutions to the gravitational field equations for both the Einstein and de Sitter models, extends his analysis to include the possibility of the curvature radius varying as a function of time.  This is achieved by replacing the constant $R$ term in equation \eqref{eqn:Einsteinstatic}, and by introducing a time-dependent scale factor \[a(t) \equiv \frac{R(t)}{R_0}\] into the metric

\begin{equation} \label{eqn:FLRW}
ds^{2}=dt^{2}-a^{2}(t)\left({\frac{{dr^{2}}}{{1-kr^{2}}}+r^{2}d\theta^{2}+r^{2}\sin^{2}\theta d\phi^{2}}\right)
\end{equation}

Then taking the Einstein field equation of GR (while noting that $ \Lambda =0$ is a valid case)
\begin{equation}
R_{\mu\nu}-\frac{1}{2}g_{\mu\nu}R=\frac{8\pi G}{c^4}T_{\mu\nu}+ \Lambda g_{\mu\nu} \label{eqn:EFE}
\end{equation}

and solving for the $x_{00}$ and $x_{ii}  (i=1,2,3)$ components of this equation with the corresponding components of the time-varying metric of equation \eqref{eqn:FLRW}, gives the well-known Friedman equations that describe the dynamics of the universe

\begin{equation}
\label{eqn:adot}
\frac{{\dot{a}^{2}}}{{a^{2}}}+\frac{k}{{a^{2}}}=\frac{8\pi G \rho}{3}
\end{equation}

\begin{equation}
\label{eqn:addot}
\frac{2\ddot{a}}{a}+\frac{\dot{a}^{2}}{a^{2}}+\frac{k}{a^{2}}=-8\pi Gp
\end{equation}
where $a$ is the cosmological scale factor, $\rho$ is matter density, and $p$ is the pressure of the cosmological fluid.  These dynamic equations imply that the universe is not static; it must either expand or contract.  However, the field equations of general relativity \eqref{eqn:EFE} do not explicitly incorporate time. The 2nd order Ricci tensor $R_{\mu\nu}$ and its derivatives employ four covariant indices in the context of GR; however, there is no implication or requirement that any of these should necessarily relate to time. Time and the concept of a dynamic universe only comes about because of the imposition of a time-dependent scale factor in Friedmann's metric. The main objection to this construction is that it imbues the FLRW metric with a spurious time dependency without proposing any physical mechanism or spacetime property that might give rise to such a time-dependent scale factor.

In the years following Friedman's seminal 1922 paper, other cosmologists published similar formulations that incorporate a time-dependent metric.  The names we now most closely associated with this work are Lem\^{a}itre \cite{Lemaitre1927}, Robertson \cite{Robertson1929}, and Walker \cite{Walker1935a}, hence the present-day identification of equation \eqref{eqn:FLRW} as the 'FLRW metric'.  However, the differential equations originally derived in Friedman's paper are still used today to describe the dynamics of the $\Lambda$CDM universe.

\section{Time in General Relativity} \label{sec:time_in_GR}

\subsection{Historical Perspective}

When Einstein formulated the general theory of relativity in the years leading up to its publication in 1915, it was only natural that he would identify time as the $x_0$ dimension, (i.e. the $x_4$ dimension in the notation used at that time), as that provided a direct link with the concept of time in the Minkowski metric of Special Relativity. The main requirement was for a second order rank-2 curvature tensor that could be equated to the corresponding stress-energy tensor in the gravitational field equation, and time was the `natural' candidate for the $x_0$  component of this tensor.  However, it should be noted as a general observation that the choice of a space-like dimension for $x_0$  still retains the essential structure of GR, provided that it remains possible to differentiate the other 3 spatial dimensions with respect to this choice of $x_0$, i.e. $x_{i,0}\ne 0$. With Einstein's choice of time as the $x_0$ dimension, it was inevitable that he should use this at the starting point for deriving the metric for his static universe,  which in turn motivated Friedman's choice of the coordinate system for his scale-varying metric.  Subsequent elaboration of the expanding universe model by Lem\^{a}itre, Robertson, Walker and other cosmologists in the first half of the 20th century understandably perpetuated this choice of dimensions in the metric.

\subsection{The advent of Quantum Gravity}

This state of affairs worked well enough until a number of theorists started to work on a program to unify GR with the more recently developed theory of quantum mechanics, with the ultimate aim of deriving an all-embracing theory of quantum gravity. At this point, it became apparent that the concept of time in GR was essentially undefined, and in any case, was incompatible with the nature of time inherent in QM's Schr\"{o}dinger equation.  One of the first attempts to address the structure of time in GR was the work of 
Arnowitt, Deser and Misner \cite{Arnowitt1959, Arnowitt1959a, Arnowitt1960, Arnowitt1960a}. Their proposal, which subsequently became known as the ADM formalism, was to treat 4D space as a series of foliations of 3D space, with the Lagrangian of each 3D slice being given by

$$\displaystyle {\mathcal {L}}=-g_{ij}\partial _{t}\pi ^{ij}-NH-N_{i}P^{i}-2\partial _{i}\left(\pi ^{ij}N_{j}-{\frac {1}{2}}\pi N^{i}+\nabla ^{i}N{\sqrt {g}}\right)$$

where the Lagrange multipliers $  N=\left(-{^{(4)}g^{00}}\right)^{-1/2}$ and $N_{i}={^{(4)}g_{0i}}$ are the shift and lapse factors that define the separation between successive foliations, respectively.   Thus, in a limited sense, $N$ and $N_i$ can be thought of imbuing a time-like ordering to the resulting 4D space.

$$\displaystyle \partial _{t}g_{ij}={\frac {2N}{\sqrt {g}}}\left(\pi _{ij}-{\tfrac {1}{2}}\pi g_{ij}\right)+N_{i;j}+N_{j;i}$$

In proposing the ADM formalism, its authors essentially decoupled GR from any explicit definition of time, and in its place treated time as an emergent property defined by the evolution of observable properties of matter (or radiation) in the universe, that is, their positions in space.  Subsequently, the ADM formalism was further developed by Baierlein, Sharp, and Wheeler \cite{Baierlein1962} to derive a more rigorous definition of time in GR as the link between successive configurations of 3D space.  Again, as with the ADM formalism, this derivation of a time label is necessary to compensate for the lack of any intrinsic definition of time within the framework of GR.  

Numerous attempts have been made to circumvent the dialectical nature of time in GR and QM.  The Hartle-Hawking no-boundary proposal \cite{Hartle1983} is a case in point.  In their model, time does not exist as a discrete dimension in the Big Bang singularity, and only emerges as the three physical dimensions expand.  Another approach to the theory of time that might apply to quantum gravity is Halliwell's histories theory \cite{Halliwell1991}.  A comprehensive review of the problem of time in relation to the formulation of a theory of quantum gravity was provided by Isham in \cite{Isham1993}, and in addition to elaborating on the multiple facets of this problem, he summarises the work undertaken by many other researchers in this field. Perhaps the status of this research effort at the time of writing can be captured by Isham's speculation: \emph{\say{Can ‘time’ still be regarded as a fundamental concept in a quantum theory of gravity, or is its status purely phenomenological? If the concept of time is not fundamental, should it be replaced by something that is: for example, the idea of a history of a system, or process, or an ordering structure that is more general than that afforded by the conventional idea of time?}}.  Interestingly, this observation is predated by a statement made by Mach in \cite{mach_2013} \emph{\say{It is utterly beyond our power to measure the changes of things by time. Quite the contrary, time is an abstraction at which we arrive through the changes of things.}}

This concept forms the basis of several recent developments in the theory of time. Two prevalent approaches are highlighted here, as propounded by Rovelli  \cite{Rovelli1991, Rovelli2008} and Barbour \cite{Barbour2001,Barbour2002}.  Both of these approaches take as their starting point the postulate that time in the context of GR is both unnecessary and undefined and, therefore, has to be treated as an emergent property of any dynamical system being studied. 

\subsection{The Rovelli proposal}
In \cite{Rovelli2008}, Rovelli summarises his approach to time as follows:

\begin{enumerate}

\item \emph{It is possible to formulate classical mechanics in a way in which the time variable is treated on equal footings with the other physical variables, and not singled out as the special independent variable. I have argued that this is the natural formalism for describing general relativistic systems.}

\item \emph{It is possible to formulate quantum mechanics in the same manner. I think that this may be the effective formalism for quantum gravity.}

\item \emph{The peculiar properties of the time variable are of thermodynamical origin, and can be captured by the thermal time hypothesis. Within quantum field theory, “time” is the Tomita flow of the statistical state $\rho$ in which the world happens to be, when described in terms of the macroscopic parameters we have chosen}

\item \emph{In order to build a quantum theory of gravity the most effective strategy is therefore to forget the notion of time all together, and to define a quantum theory capable of predicting the possible correlations between partial observables.}

\end{enumerate}

In \cite{Rovelli2017} Rovelli discusses the concept of eternalism, also known as the \emph{block universe}, which was originally propounded by Einstein and which views spacetime as a block in which the past, present, and future co-exist on an equal footing.

\subsection{The Barbour-Bertotti-Jacobi approach}
Barbour's theory, which was initially developed in collaboration with Bertotti \cite{Barbour1982},  is essentially an extension of the concept of ephemeris time, commonly employed by the astronomical community, to include all objects in the universe.  Starting with the requirement for the conservation of the total energy of the universe
$$  E = V + T$$
where $V$ is the gravitational potential energy of the system and the kinetic energy $T$ is given by
$$ T = \sum_i{\frac{m_i}{2}\left(\frac{\delta d_i}{dt}\right)^2}$$

then the ephemeris time defined by this system is 

$$ \delta t =  \sqrt{\frac{\sum_i{m_i\left(\delta d_i\right)^2}}{2(E-V)}} $$

Thus it is possible to define a time-like coordinate purely in terms of changes in the configuration of matter in the universe.  

This initial study was further developed by Barbour in collaboration with O'Murchadha, Anderson, Foster  and Kelleher \cite{Barbour2002, Anderson2003,Anderson2005, Barbour2009}.  Subsequently, Anderson has continued to pursue research into the nature of time in GR \cite{Anderson2010, Anderson2012a}, and has published a comprehensive review article summarising the broad landscape of research into the problem of time  \cite{Anderson2012}.

Research into the nature of time in GR and the problem of time in quantum gravity continues to this day, for example, the suggestion by Magueijo and Smolin \cite{Magueijo2019} that the universe, at least from a quantum perspective, may be decoupled from time.

\subsection{Conclusion}

Our aim in this section has not been to elaborate on potential solutions to the problem of time in quantum gravity or general relativity, but merely to highlight the fact that GR does not inherently embody any notion of time.  Hence, this leaves open the possibility of constructing a solution to the gravitational field equations of GR that explicitly excludes a time dimension.  The possibility is explored in greater depth in section \ref{sec:exochronous}.

\section{A Timeless Metric  \label{sec:exochronous}}

\subsection{Motivation for an alternative metric}

In section \ref{sec:problems} we saw how a cosmological model incorporating the FLRW metric requires the presence of two unidentified fields or particles - dark energy and dark matter - in order for the differential equations that govern the dynamics of the universe to generate  results that agree with astronomical observations.  Furthermore, a metric describing the behaviour of the universe on the largest scales that incorporates a time-varying scale factor raises numerous physical and philosophical issues that call into question its fitness for purpose.  At the same time, we should not lose sight of the fact that GR is a remarkably successful theory of gravitation when allied with the appropriate metric.  The Schwarzschild solution to the Einstein field equations provides an accurate description of the classical tests of GR at solar system scales, including the precession of Mercury's orbital perihelion, deflection of light by the sun, photon gravitational redshift, and gravitational time dilation. (See \cite{Will2014} for a detailed review of these GR tests). 

Essentially, what we are seeking to achieve with this current exercise is to ascertain whether it is, in principle, possible to formulate an alternative metric that describes the large-scale behaviour of the universe as a whole, but which does not suffer from the drawbacks identified in section \ref{sec:problems}.  Some possible objectives for any alternative metric include the following attributes:
\begin{itemize}
    \item scale free, or at least scale invariant
    \item giving rise to a static solution to the gravitational field equations, while remaining compatible with the observation of the Hubble redshift
    \item adhering to the Cosmological Principle, by being both homogeneous and isotropic
    \item able to retain particle histories, without necessarily invoking a time-like dimension.
\end{itemize}

Implicit in these objectives is the implication that any metric describing a static universe will \emph{not} lead to any equivalence to the Friedman equations, and hence will render the concepts of dark matter and dark energy irrelevant, at least in terms of their impact on large-scale cosmic dynamics.  This in turn implies that some other mechanism independent of the metric is required to explain the observed cosmological redshift-distance relation.

\subsection{The Exochronous Metric}

The objective in this section is to construct an alternative metric that replaces the time dimension used in the Einstein and FLRW metrics with a space-like ordering dimension. The new metric must, of course, be compatible with solutions to Einstein's gravitational field equations.  We begin with the most generalised form for representing a curved space as a 3-sphere $\mathcal{S}^3$ embedded in four-dimensional Euclidean space $\mathcal{E}^4$, such that

\[R^2=w^2+x^2+y^2+z^2\]
with the line element given by
\[d\sigma^2 =|(dw, dx, dy, dz)|^2 = R^2[d\chi^2+\sin^2\chi(d\theta^2 + \sin^2\theta d\phi^2)]\]
where $R$ is the radius of the 3-sphere in spherical space.

Transforming to elliptical space, where $r=\sin\chi$ gives
\begin{equation} \label{eqn:metric}
	d\sigma^2= R^2\left(\frac{dr^2}{1-kr^2} + r^2(d\theta^2+\sin^2\theta d\phi)\right)
\end{equation}

This describes a homogeneous, isotropic 3D space of curvature $1/R^2$, where $k = -1, 0$, or $+1$. 

Thus far, this has followed the same process used to construct the Einstein and FLRW metrics.  However, we now divide the $x_0$ component (=d$R$) into $N(R)$ foliations, such that $N$ is itself a function of $R$.

From which 

\begin{equation} \label{eqn:exochronous}
ds^{2}= \frac{\mathrm{d}R^2}{N(R)} + R^{2}\left({\frac{{\mathrm{d}r^{2}}}{{1-kr^{2}}}+r^{2}\mathrm{d}\theta^{2}+r^{2}\sin^{2}\theta \mathrm{d}\phi^{2}}\right)
\end{equation}

With this timeless or exochronous metric, we can now construct the corresponding Ricci and Einstein tensors, and hence solve the gravitational field equations, as shown in \ref{app:exo}.

\subsection{Solutions to the Einstein Field Equations}
From  \ref{app:exo}

$x_0$ component:

\begin{equation}
    \frac{3N}{R^{2}}  = \kappa \rho 
     \label{eqn:exo4}
\end{equation}
where $\kappa \equiv \frac{8  \pi G}{c^4}$ is Einstein's gravitational constant.

$x_i$ component (i = 1,2,3):

\begin{equation}
    \frac{1}{R}\frac{dN}{dR} + \frac{N}{R^2} = \kappa p
\end{equation}

For a universe in which the pressure term $p$ is effectively zero we have

\begin{equation}
   \frac{dN}{dR} = - \frac{N}{R} 
\end{equation}

which has the solution
\begin{equation} \label{eqn:exo6}
     N = \alpha/R 
\end{equation}
where $\alpha$ is a constant of integration.

We can consider \eqref{eqn:exo4} to define the Riemann curvature for a single hypersurface with foliation index $N$, thus 

\begin{equation}
    \frac{1}{R_N^{2}}  = \frac{1}{N}\frac{\kappa \rho }{3} 
\end{equation}

\begin{figure}[ht]
\includegraphics[scale =0.5]{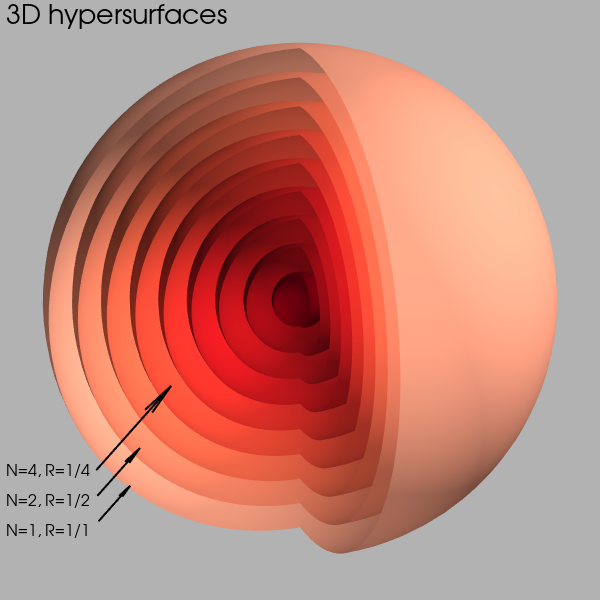}
\caption{Metric Foliations \label{fig:shells}}
\end{figure}

If the average energy density $\rho$ is constant, this  equation describes a static universe with a single 3D hypersurface (or foliation) of constant curvature $1/R^2$. However, the Ricci tensor derived from the Exochronous metric is capable of encoding the curvatures resulting from multiple foliations generated by the evolution of the stress-energy tensor, $T_{\mu \nu}$. Thus we can construct an expression for the cumulative curvature resulting from the entire ensemble of hypersurfaces

\begin{equation} \label{eqn:SigmaGR}
    \frac{1}{R_{\Sigma}^{2}} = \sum_{N_1}^{N_0} \frac{1}{N}\frac{\kappa \rho }{3} 
\end{equation}

where $N_0$ is the present-day foliation count and $N_1$ is the initial foliation count. 
This can be visualised as a series of concentric 3D hypersurface shells or foliations, as illustrated in Figure \ref{fig:shells}.  We employ the term $\Sigma$GR to refer to the model defined by \eqref{eqn:SigmaGR} in the remainder of this study.

It can be seen from \eqref{eqn:SigmaGR} that this ensemble of metric foliations is essentially the well-known harmonic series 

\begin{equation}
    \sum_{k=1}^{N} \frac{1}{k} = 1 + \frac{1}{2} + \frac{1}{3}  + \frac{1}{4} \cdots  + \frac{1}{N} 
\end{equation}

Even though the foliation index $N$ is inherently quantised, we can legitimately approximate this sum as an integral \footnote{In fact it can be shown that $\sum_{k=1}^{n} {\frac{1}{k}}  = \log{n} + \gamma$, where $\gamma \simeq 0.5772$ is Euler's constant}, thus

\begin{align}
    \frac{1}{R_{\Sigma}^{2}}  &= \frac{\kappa \rho }{3} \int_{N_1}^{N_0} \frac{1}{N} \mathrm{d}N \label{eqn:sigmaGR}\\ 
    &= \frac{\kappa \rho }{3}  \Big [ \log{N} \Big ]_{N_1}^{N_0} \\
    &= \frac{\beta \kappa \rho }{3}  \label{eqn:exo5}
\end{align}

where
\begin{equation} \label{eqn:beta1}
    \beta = \log({N_0}/{N_1})
\end{equation}

\subsection{Analysis \label{sec:analysis}}

From \eqref{eqn:exo5} we see that while the universe is evidently static, in that it does not re-scale over time, it \emph{does} evolve in the sense that its cumulative Riemann curvature increases with the proliferation of the foliations created by the matter fields that comprise the energy content of the universe.  It is therefore instructive to quantify the size of the $\beta$ factor in  \eqref{eqn:exo5}, using the values applicable to the primary matter content of the universe, that is, baryons (assuming one ignores the existence of dark matter).  However, as it stands the formulation of equation \eqref{eqn:sigmaGR} is over-simplistic in that it assumes a continuous and progressive evolution of the foliation index from $N=1$ to the present-day.  A more realistic scenario (which will be expounded in greater detail in a forthcoming article) is a two-phase scenario.  The first phase involves an exponential doubling in baryon number up to an inflection point where the gravitational force acting on a baryon is in equilibrium with the electromagnetic force, at which point the system enters a second phase of linear evolution. (This is analogous to the transition between an inflationary phase and the subsequent slower evolution of the cosmic scale factor in the standard $\Lambda$CDM model.)  We must therefore expand equation \eqref{eqn:sigmaGR} to take into account the cumulative contributions of each phase to the total curvature, thus
\begin{align}
    \frac{1}{R_{\Sigma}^{2}}  &= \frac{\kappa \rho }{3} \left[ \int_{N_1}^{N_2} \frac{1}{N} dN +  \int_{N_3}^{N_0} \frac{1}{N} dN \right]  \label{eqn:sigmaGR2}\\ 
    &= \frac{\kappa \rho }{3} \left( \beta_1 + \beta_2 \right)  = \frac{\beta \kappa \rho }{3} 
\end{align}

where 
\begin{align}
    \beta_1 &= \log({N_2}/{N_1}) \\
    \beta_2 &= \log({N_0}/{N_3}) \label{eqn:beta2}\\
    \beta &= \beta_1 + \beta_2 \label{eqn:beta+}
\end{align}

Now we must determine the values for $N_2$, $N_3$, and $N_0$ that are used to calculate $\beta$.  To estimate $N_2$, the foliation index at the end of the baryogenesis epoch, we first need to establish the total baryon number, $n$, in the universe.  
\begin{equation}
    n = \frac{4\pi}{3}\left(\frac{c}{H_0} \right)^3 \frac{\rho_b}{m_p}
\end{equation}

where $c/H_0$ is an approximation for the radius of the universe, $\rho_b$ is the baryonic matter density, and $m_p$ the proton mass. Using appropriate values for $H_0$ and $\rho_b$ from \cite{Planck2015}  gives a value for $n \simeq 1.13 \times 10^{78}$ baryons.  Assuming that each bifurcation of the baryon field gives rise to a metric foliation then we have a foliation count at the end of the baryogenesis epoch given by $N_2 = \log_2(n) \simeq 259$.  This gives a value for $\beta_1 = \log(N_2/N_1) \simeq 5.56 $.

Turning to the second evolutionary phase, commencing at the end of the baryogenesis epoch, we have from \eqref{eqn:exo6}

\begin{equation} \label{eqn:N0}
     N_0 \propto \lambda_0^{-1}
\end{equation}

where $\lambda_0$ is the present day proton Compton wavelength, and similarly

\begin{equation} \label{eqn:N3}
     N_3 \propto  \lambda_3^{-1}
\end{equation}

where $\lambda_3$ is the proton Compton wavelength at the end of the epoch of baryogenesis, at which time it is assumed that baryons occupy the entire spatial volume of the universe. We can determine an approximate value for $\lambda_3$ from the present-day measured baryonic matter density of the universe, $\rho_b$

\begin{equation} \label{eqn:lambda3}
    \lambda_3 = \left( \frac{4 \pi \rho_b }{3 m_p} \right)^{-1/3}
\end{equation}

where $m_p$ is the proton rest mass.

Combining \eqref{eqn:beta2}, \eqref{eqn:N0}, \eqref{eqn:N3}, and \eqref{eqn:lambda3} yields

\begin{equation} \label{eqn:beta0}
    \beta_2 = \log \left[\frac{( 3 m_p / 4 \pi \rho_b )^{1/3}}{\lambda_0} \right]
\end{equation}

Using the value of $\Omega_b = 0.0486$ from \cite{Planck2015}, the standard values for $m_p$ and $\lambda_0$ give $\beta_2 \simeq 34.5$. Thus, from \eqref{eqn:beta+}
\begin{equation}  \label{eqn:Beta}
    \boxed{\beta\simeq 40.0} 
\end{equation}

\subsection{Critical energy density \label{sec:rho_crit}}

The $\Lambda$CDM model introduces the concept of a critical energy density, where the energy density required for the curvature term in \eqref{eqn:adot} is zero, which leads to 

\begin{eqnarray}\label{eqn:rho_crit}
    \left( \frac{\dot{a}}{a} \right)^2 = \frac{8 \pi G \rho_{\mathrm{crit}}}{3} \nonumber  \\
    \rho_{\mathrm{crit}} = \frac{3 H_0^2}{8 \pi G}
\end{eqnarray}

where $H_0 \equiv \dot{a}/a $ defines the present day-value of the Hubble parameter.  However, it is well known that the dynamical equation of (\ref{eqn:adot}) can be derived without the use of the full GR formalism incorporating the FLRW metric, using only the application of the laws of energy conservation in a Newtonian universe. Hence, the concept of a critical energy density should still be applicable to the $\Sigma$GR  model, even though it has been derived from the Exochronous metric rather than the FLRW metric. In this case we must apply the present day value of $\beta$, as defined in \eqref{eqn:beta+}, such that

\begin{equation}\label{eqn:rho_crit_beta}
        \rho_{\mathrm{crit}} = \frac{3 H_0^2}{8 \pi G \beta}
\end{equation}

Using the appropriate values for $H_0$ from \cite{Planck2015} and $\beta$ from \eqref{eqn:Beta} gives 
\begin{equation}
    \boxed{ \rho_{\mathrm{crit}} = 2.15 \times 10^{-28} \; kg/m^3}
\end{equation} 

which is within a factor of $2$ of the baryonic matter density ($\rho_{\mathrm{b}} = 4.19 \times 10^{-28} \; kg/m^3$) derived from early universe calculations in the $\Lambda$CDM model.  This  naturally leads to the speculation that, with a more detailed understanding of the thermodynamic environment in the primordial universe in a $\Sigma$GR model, we may find $\rho_{\mathrm{crit}} = \rho_b$.  In other words, the measured cosmic baryon matter density may be sufficient on its own to give rise to a closed universe in the $\Sigma$GR model without invoking any additional dark energy or dark matter content.

\subsection{The Newtonian Gravitational Constant \label{sec:NewtonianG}}

Taking the observation in section \ref{sec:rho_crit}, that $\rho_{\mathrm{crit}} = \rho_b$, to its logical conclusion, we can rewrite equation \eqref{eqn:rho_crit_beta} as
\begin{equation}\label{eqn:G}
        G = \frac{3 H_0^2}{8 \pi  \beta \bar{\rho}}  = \frac{c^3}{2 H_0  M \beta}
\end{equation}
where $\bar{\rho}$ is the mean matter density of the universe, $M$ is the total mass of the universe and $c/H_0=R$ represents the `radius' of the universe in present-day units.  Thus we see that in the $\Sigma$GM model, the gravitational `constant', $G$, that appears in Newton's law of gravitation and in Einstein's field equations of gravity is in fact composed of two components: 

\begin{enumerate}
    \item a unit conversion factor that converts between standard atomic based units of mass and length, and the equivalent `natural' cosmological units (where $M=R=1$), and is constant provided that the scale factor defined by the two length scales remains constant. 
    \item an evolving factor, $\beta$, caused by the accumulation of curvature in the Einstein curvature tensor, $G_{\mu\nu}$, which increases logarithmically with the evolution of the universe.
\end{enumerate}

\section{Discussion and Conclusions \label{sec:conclusions}}

In this article, it has been suggested that the $\Lambda$CDM model, which forms the cornerstone of the current cosmological paradigm is in many ways deeply flawed.  We have argued that the use of the FLRW metric, incorporating a time-varying scale factor, is not appropriate for solving Einstein's gravitational field equations for the universe as a whole.  In its place we have proposed an alternative, the Exochronous metric, in which the time dimension is replaced by a hyperspatial `ordering' dimension, and we have shown how this leads to a static solution to the gravitational field equations of General Relativity.  Furthermore, we described how this new solution permits the summation of successive foliations of the energy/matter fields to generate a cumulative curvature field encoded in the Ricci curvature tensor, a process that we denote as the $\Sigma$GR formalism.

Inevitably, such a drastic extension of the Einstein field equations of GR will have a profound impact in many areas of cosmology.  Some of the principal implications are presented in the following sections.

\subsection{Radiation energy density in a static universe}
One of the most striking consequences of a static universe model is that the photon-baryon energy ratio established by processes occurring in the early universe will remain largely unchanged as the universe continues to evolve, since the loss of photon energy due to the wavelength increase in an expanding universe will not be a factor.  Measurements of the present day photon flux in the CMB indicate that individual photon energies are consistent with a $2.7$K thermal background, implying that the photon number count has increased over time as the universe has cooled.  The only plausible mechanism that could give rise to this effect is that of double-Compton scattering, whereby a high energy photon interacts with a free electron to produce two photons of lower energy.  This process would only be applicable to epochs in which hydrogen is ionized, which would be the case in the universe prior to the epoch of recombination, and potentially also in the low redshift universe after the epoch of re-ionization.  

\subsection{Impact on observed galaxy angular size to distance relationship}
 The expanding universe described by the $\Lambda$CDM model predicts that the angular diameter distance of an object from a present day observer will increase as a function of redshift up to an inflection point at $z \simeq1.6$, after which it will start to decrease with increasing redshift. Thus the observed angular diameter of a galaxy of a given size will appear to decrease with redshift up to this inflection point, and then start to increase again as its image is `stretched' by the Hubble expansion of the universe. In a static universe, such as that predicted by the $\Sigma$GR model, no such stretching takes place and hence the observed angular diameter of galaxies should appear progressively smaller throughout the whole observable redshift range.  There is already some evidence for this effect from HST surveys (see \cite{Bridge2019}, for example), but we can expect increasing evidence for this anomalous angular size-distance relationship to emerge from the observations of high redshift galaxies made possible by the forthcoming JWST mission \cite{Steinhardt2021}. 

\subsection{An explanation for Dark Matter}
The $\Sigma$GR model potentially holds out the possibility of providing an alternative explanation for the phenomenon of dark matter that is of a non-particulate nature.  With the replacement of time in the metric with a continuous space-like dimension, it is possible to envision how the curvature of the metric induced by matter (and radiation) might be cumulative as the universe evolves, as illustrated by the foliations in Figure \ref{fig:shells}.  In section \ref{sec:rho_crit}, we showed that this cumulative curvature effect is sufficient, when combined with the observed baryonic matter density, to give rise to a closed universe without the need for any additional dark matter.  This cumulative curvature mechanism can, in principle, be applied to any accumulation of matter, such as a galaxy. Over the course of its evolution from a primordial gas cloud a galaxy could accumulate a curvature `halo', which might reproduce the gravitational effects (namely, flat rotation curves) currently ascribed to dark matter.

\subsection {Accounting for the Hubble redshift}
At first sight, it might appear that the concept of a static universe is a considerable backward step compared to the status quo, which, despite the many issues outlined in section \ref{sec:problems}, does at least result in cosmic dynamics that are largely in accord with observations.  As a minimum requirement, any alternative cosmological model must be able to explain the observed Hubble redshift. By removing this function from the metric, and if the concept of `tired light' is excluded, we are left with only one viable alternative explanation: the observed evolution in the redshift (or scale factor) must be due to the progressive contraction of atomic matter over cosmological timescales.  One of the challenges that will need to be addressed by a fully developed cosmological model based on this principle is to reproduce the effects of deceleration and acceleration on the growth rate of the cosmic scale factor that is currently ascribed to dark matter and dark energy within the Friedman equations of the $\Lambda$CDM model. Hopefully the application of the $\Sigma$GR formalism to the evolution of a baryon's gravitational potential field, coupled with the time-dependent Newtonian gravitational `constant' described in section \ref{sec:NewtonianG}, has the potential to resolve this puzzle.

\subsection{Implications for the early universe}
In principle one could adopt a static universe model while retaining many of the features of the standard HBB-$\Lambda$CDM model.  However, some of these components, such as inflation, do not sit easily within the static universe model.  Nevertheless, there are certain processes that must occur in the early universe if the astrophysical ingredients we observe today are to be accounted for, albeit with the possibility that their detailed workings differ markedly from their equivalents in the HBB model.  Some of these processes that require further elucidation include the following:

\begin{itemize}
    \item Baryogenesis - exponential growth in baryon number in a rapidly evolving gravitational field as a replacement for inflation
    \item Primordial nucleosynthesis - as a stochastic exothermic process, giving rise to Gaussian perturbations in primordial matter density, and $\eta_\gamma  \simeq 1$
    \item Baryon acoustic oscillations - evolution of the speed of sound in the early universe with an evolving $\eta_\gamma$, and the impact on BAO $r_{\mathrm{drag}}$ scale
\end{itemize}

\subsection{Varying Newtonian gravitational `constant'}
As described in section \ref{sec:NewtonianG}, the $\Sigma$GR model replaces the constant Newtonian $G$ with an evolving cumulative curvature factor $\beta$ (and a constant unit conversion factor). While the logarithmic nature of $\beta$ means that its present-day growth rate is very small, the fact that $G$ is no longer a fundamental constant has profound implications for the study of quantum gravity.  Without a constant $G$, the concept of the Planck scale is no longer applicable:

Planck Length:    $ L_P  = \sqrt {\frac{{\hbar G}}{{c^3 }}} \simeq 10^{-35}m $

Planck Time:  $  T_P  = \frac{{L_P }}{c} \simeq 10^{-43}s $

Planck Mass: $  M_P = \sqrt {\frac{{\hbar c}}{G}} \simeq 10^{-8}kg $

Indeed, it is plausible to speculate that the $\Sigma$GM model is in itself a component of a viable theory of quantum gravity, in that it incorporates the harmonic series of hyperspatial foliations, which are themselves quantised.

Another consequence of a varying $G$ is that the integral that defines the cumulative curvature factor $\beta$ will vary not only as a function of the universe's evolution but will also be dependent on the location in 3D space.  In particular, the value of $\beta$ in the proximity of a massive object, such as a black hole, is much lower than its value in free space. This leads to the possibility that, rather than having a central singularity, which is the inevitable outcome in standard GR, all the black hole's mass is located at the event horizon, which then becomes an impenetrable boundary in $\Sigma$GR.

\subsection{Conclusion}
The main objective of this exercise was to construct a cosmological model that, by doing away with time, has the potential to explain a number of observed features of our universe without the need to invoke any undiscovered dark ingredients. The $\Sigma$GR model we have described holds out the tantalising prospect of being able to shed light on several other areas of cosmology, and some of the consequences identified in this section will be explored further in subsequent articles.

\appendix

\section{Solving the Einstein Field Equations \label{app:metric}}
\subsection{Einstein Metric}  \label{app:Einstein_metric}

The line element is defined as

\begin{equation} \label{eqn:FLRW1}
ds^{2}=-c^2dt^{2}-R^{2}\left({\frac{{dr^{2}}}{{1-kr^{2}}}+r^{2}d\theta^{2}+r^{2}\sin^{2}\theta d\phi^{2}}\right)
\end{equation}

which gives non-zero diagonal metric elements 
\[ g_{00} = -1; g_{11} =  \frac {R^2}{1-kr^2}; g_{22} = R^2r^2; g_{33} = R^2r^2\sin^2 \theta; \]

Now we calculate the Christoffel symbols using

\[\Gamma^{\sigma}_{\mu \nu} = \frac{1}{2}g^{\sigma \tau}(g_{\mu \tau, \nu} + g_{\tau \nu, \mu} - g_{\mu \nu, \tau}) \]

which leads to the Einstein tensor:

\[\left[\begin{matrix}-\frac{3 c^{2}}{R^{2}} & 0 & 0 & 0\\0 & \frac{1}{r^{2} - 1} & 0 & 0\\0 & 0 & - r^{2} & 0\\0 & 0 & 0 & - r^{2} \sin^{2}{\left (\theta \right )}\end{matrix}\right] \]

Solutions:

Timelike components, $i=0$

$$G_{00}g^{00} = \dfrac{8 \pi}{c^4}T_{00}g^{00}$$

\begin{equation} \label{eqn:Einstein1} 
 \frac{1}{R^{2}} = \dfrac{8 \pi G \rho}{3c^2} 
\end{equation} 

Spatial components, $i = 1,2,3$

\[ G_{ii}g^{ii} = \dfrac{8 \pi}{c^4}T_{ii}g^{ii} \]

\begin{equation}\label{eqn:Einstein2}
     \frac{1}{R^{2}} = \dfrac{8 \pi G p}{c^2}
\end{equation}

If $p=0$, as it would be in a low-density region of the universe, then we have $R = \infty$ from \eqref{eqn:Einstein2}, which is inconsistent with \eqref{eqn:Einstein1}.

Einstein's proposed solution to this problem was the addition of a cosmological constant $\lambda$ to the gravitational field equation, so that \eqref{eqn:Einstein2} now becomes

\[ G_{ii}g^{ii} + \lambda g_{ii} = \dfrac{8 \pi}{c^4}T_{ii}g^{ii} \]

which gives the solution 
\begin{equation}
         \frac{1}{R^{2}} = \lambda
\end{equation}

And substituting back into \eqref{eqn:Einstein1}.

\begin{equation}
   \lambda = \frac{4 \pi G \rho}{c^2}
\end{equation}


\subsection{Exochronous Metric \label{app:exo}}

The line element is defined as

\begin{equation} \label{eqn:exochronous1}
ds^{2}= \frac{\mathrm{d}R^2}{N(R)}  + R^{2}\left({\frac{{\mathrm{d}r^{2}}}{{1-kr^{2}}}+r^{2}\mathrm{d}\theta^{2}+r^{2}\sin^{2}\theta \mathrm{d}\phi^{2}}\right)
\end{equation}

which gives non-zero diagonal metric elements 
\[ g_{00} = \frac{1}{N(R)}; g_{11} =  -\frac {R^2}{1-kr^2}; g_{22} = -R^2r^2; g_{33} = -R^2r^2\sin^2 \theta; \]

Next we calculate Christoffel symbols using

\[\Gamma^{\sigma}_{\mu \nu} = \frac{1}{2}g^{\sigma \tau}(g_{\mu \tau, \nu} + g_{\tau \nu, \mu} - g_{\mu \nu, \tau}) \]

which results in the following non-zero elements:
\begin{align*}
& \Gamma^{R}_{RR}  = \frac{{\frac{\mathrm{d}N}{\mathrm{d}R}}}{2N^2} \\
&\Gamma^{R}_{rr}  =  -R  \\
& \Gamma^{R}_{\theta \theta} = \Gamma^{r}_{\theta \theta}  = -R r^2  \\
& \Gamma^{R}_{\phi \phi}  =\Gamma^{r}_{\phi \phi} =  R r^2 \sin^2 \theta \\
%
& \Gamma^{\theta}_{r \theta}  = \Gamma^{\theta}_{\theta r} =\Gamma^{\phi}_{r \phi} =\Gamma^{\phi}_{\phi r}  = R^2 r \\
&\Gamma^{\theta}_{\phi \phi}  = - \frac{R^{2} r^{2} \sin{\left(2 \theta \right)}}{2} \\
& \Gamma^{R}_{\phi \phi}  =\Gamma^{r}_{\phi \phi} =  R^2 r^2 \sin^2 \theta \\
& \Gamma^{R}_{\theta \theta} = \Gamma^{r}_{\theta \theta}  = R   \\
\end{align*}

These can be used to construct the Ricci curvature tensor

\[ \displaystyle \left[\begin{matrix}- \frac{3N^\prime}{2 R N} & 0 & 0 & 0\\0 & - \frac{RN^\prime+ 4 N}{2}  & 0 & 0\\0 & 0 & - \frac{r^{2} \left(RN^\prime + 4 N\right)}{2} & 0\\0 & 0 & 0 & - \frac{r^{2} \left(RN^\prime+ 4 N\right) \sin^{2}{\left(\theta \right)}}{2}\end{matrix}\right] \]

where $N^\prime \equiv \frac{\mathrm{d} N}{\mathrm{d} R}$ and $N \equiv N(R)$,

from which can be obtained the Ricci scalar
\[\displaystyle - \frac{3 R N^\prime + 6 N}{R^{2}}\]

These combine to give the Einstein tensor

\[ G_{\mu \nu} = R_{\mu \nu} - \frac{1}{2}R g_{\mu \nu} = \]

\[ \displaystyle \left[\begin{matrix}\frac{3}{R^{2}} & 0 & 0 & 0\\0 & R N^\prime + N & 0 & 0\\0 & 0 & r^{2} \left(R N^\prime + N \right) & 0\\0 & 0 & 0 & r^{2} \left(R N^\prime + N\right) \sin^{2}{\left(\theta \right)}\end{matrix}\right] \]

Solutions:

$x_{00}$ components

\[G_{00}g^{00} = \dfrac{8 \pi}{c^4}T_{00}g^{00}\]

\begin{equation} \label{eqn:exo1}
 \frac{1}{R^{2}} = \frac{1}{N}\dfrac{8 \pi G \rho}{3c^2} 
\end{equation} 

$x_{ii}$ components  (i = 1,2,3)

\[ G_{ii}g^{ii} = \dfrac{8 \pi}{c^4}T_{ii}g^{ii} \]

which with pressure terms $T_{ii} =0 $ gives 
\[
   \frac{dN}{dR} = - \frac{N}{R} 
\]

which has the solution 

\begin{equation}\label{eqn:exo2}
    N = \alpha/R 
\end{equation}

where  $\alpha$ is a constant of integration

Combining equations \eqref{eqn:exo1} and \eqref{eqn:exo2}
\[ \frac{\alpha}{R^{3}} = \dfrac{8 \pi G \rho}{3c^2} \]

\begin{equation}\label{eqn:exo3}
      R= \left( \dfrac{3 \alpha c^2} {8 \pi G \rho} \right)^{1/3}
\end{equation}

\bibliographystyle{plain}
\bibliography{references}

\begin{thebibliography}{10}

\bibitem{Anderson2012}
E.~Anderson.
\newblock {Problem of time in quantum gravity}.
\newblock {\em Annalen der Physik}, 524(12):757--786, 2012.

\bibitem{Anderson2005}
E.~Anderson, J.~Barbour, B.~Z. Foster, B.~Kelleher, and N.~Ó Murchadha.
\newblock {The physical gravitational degrees of freedom}.
\newblock {\em Classical and Quantum Gravity}, 22(9):1795--1802, 2005.

\bibitem{Anderson2010}
Edward Anderson.
\newblock {Scaled Triangleland Model of Quantum Cosmology}.
\newblock {\em arXiv:1005.2507}, 2010.

\bibitem{Anderson2012a}
Edward Anderson.
\newblock {Machian Time Is To Be Abstracted From What Change?}
\newblock {\em arXiv:1209.1266}, 2012.

\bibitem{Anderson2003}
Edward Anderson, Julian Barbour, and Brendan Foster.
\newblock {Scale-invariant gravity : geometrodynamics}.
\newblock {\em Classical and Quantum Gravity}, 20:1571, 2003.

\bibitem{Arnowitt1959}
R.~Arnowitt and S.~Deser.
\newblock {Quantum Theory of Gravitation: General Formulation and Linearized
  Theory}.
\newblock {\em Physical Review}, 113(2), 1959.

\bibitem{Arnowitt1960}
R.~Arnowitt, S.~Deser, and C.~W. Misner.
\newblock {Canonical variables for general relativity}.
\newblock {\em Physical Review}, 117(6):1595--1602, 1960.

\bibitem{Arnowitt1960a}
R.~Arnowitt, S.~Deser, and C.~W. Misner.
\newblock {Energy, and the criteria for radiation in general relativity}.
\newblock {\em Physical Review}, 118(4), 1960.

\bibitem{Arnowitt1959a}
R.~Arnowitt, S.~Deser, and C.W. Misner.
\newblock {Dynamical Structure and Definition of Energy in General Relativity}.
\newblock {\em Physical Review}, 116, 1959.

\bibitem{Baierlein1962}
Ralph~F. Baierlein, David~H. Sharp, and John~A. Wheeler.
\newblock {Three-Dimensional Geometry as Carrier of Information about Time}.
\newblock {\em Physical Review}, 126(5), 1962.

\bibitem{Barbour2009}
J.~Barbour.
\newblock {The Nature of Time}.
\newblock {\em arXiv preprint gr-qc/0903.3489}, 2009.

\bibitem{Barbour2001}
Julian Barbour.
\newblock {\em {The end of time: The next revolution in physics}}.
\newblock Oxford University Press, 2001.

\bibitem{Barbour2002}
Julian Barbour, Brendan Foster, and Niall~Ó Murchadha.
\newblock {Relativity without relativity}.
\newblock {\em Classical and Quantum Gravity}, 19:3217--3248, 2002.

\bibitem{Barbour1982}
Julian~B. Barbour and B.~Bertotti.
\newblock {Mach's principle and the structure of dynamical theories}.
\newblock {\em Proceedings of the Royal Society of London, Serie A},
  382:295--306, 1982.

\bibitem{Basset2010}
Bruce~A. Basset and Renee Hlozek.
\newblock {Baryonic acoustic oscillations}.
\newblock In {\em Dark Energy}. Cambridge University Press, 2010.

\bibitem{Bernal2016}
José~Luis Bernal, Licia Verde, and Adam~G. Riess.
\newblock {The trouble with H0}.
\newblock {\em Journal of Cosmology and Astroparticle Physics}, 2016(10), 2016.

\bibitem{Bertone2005}
Gianfranco Bertone, Dan Hooper, and Joseph Silk.
\newblock {Particle dark matter: Evidence, candidates and constraints}.
\newblock {\em Physics Reports}, 405(5-6):279--390, 2005.

\bibitem{Bridge2019}
Joanna~S. Bridge, Benne~W. Holwerda, Mauro Stefanon, Rychard~J. Bouwens,
  Pascal~A. Oesch, Michele Trenti, Stephanie~R. Bernard, Larry~D. Bradley,
  Garth~D. Illingworth, Samir Kusmic, Dan Magee, Takahiro Morishita, Guido~W.
  Roberts-Borsani, Renske Smit, and Rebecca~L. Steele.
\newblock { The Super Eight Galaxies: Properties of a Sample of Very Bright
  Galaxies at 7 < z < 8 }.
\newblock {\em The Astrophysical Journal}, 882(1):42, 8 2019.

\bibitem{Carr2016}
Bernard Carr, Florian K{\"{u}}hnel, and Marit Sandstad.
\newblock {Primordial black holes as dark matter}.
\newblock {\em Phys. Rev. D}, 94(8):83504, 10 2016.

\bibitem{DeSitter1917}
W.~de~Sitter.
\newblock {On the relativity of inertia. Remarks concerning Einstein's latest
  hypothesis}.
\newblock {\em Amsterdam Proc.}, 19(September):1217--1225, 1917.

\bibitem{Durrer2008}
Ruth Durrer and Roy Maartens.
\newblock {Dark energy and dark gravity: Theory overview}.
\newblock {\em General Relativity and Gravitation}, 40(2-3):301--328, 2008.

\bibitem{Efstathiou1990}
G~Efstathiou, W~J Sutherland, and S~J Maddox.
\newblock {The cosmological constant and cold dark matter}.
\newblock {\em Nature}, 348(6303):705--707, 1990.

\bibitem{Einstein1917}
A~Einstein.
\newblock {Cosmological considerations in the general theory of relativity}.
\newblock {\em Sitz. K{\"{o}}nig. Preuss. Akad.}, pages 142--152, 1917.

\bibitem{Friedman1999}
A~Friedman.
\newblock {On the Curvature of Space}.
\newblock {\em Gen. Rel. Grav.}, 31(12):1991--2000, 1999.

\bibitem{Guth1981}
Alan~H. Guth.
\newblock {Inflationary universe: A possible solution to the horizon and
  flatness problems}.
\newblock {\em Physical Review D}, 23(2):347--356, 1981.

\bibitem{Halliwell1991}
Jonathan~J. Halliwell.
\newblock {Global spacetime symmetries in the functional Schr{\"{o}}dinger
  picture}.
\newblock {\em Physical Review D}, 43(8):2590--2609, 1991.

\bibitem{Hartle1983}
J.B. Hartle and S.W. Hawking.
\newblock {Wave function of the Universe}.
\newblock {\em Physical review D}, 28(12):2060--2975, 1983.

\bibitem{Isham1993}
C.~J. Isham.
\newblock {Canonical Quantum Gravity and the Problem of Time}.
\newblock {\em Integrable Systems, Quantum Groups, and Quantum Field Theories},
  pages 157--287, 1993.

\bibitem{Kutschera2006}
M.~Kutschera and M.~Dyrda.
\newblock {Coincidence of Universe age in LambdaCDM and Milne cosmologies}.
\newblock {\em arXiv:astro-ph/0605175}, 2006.

\bibitem{Lemaitre1927}
G.~Lema{\^{i}}tre.
\newblock {Un Univers homog{\`{e}}ne de masse constante et de rayon croissant
  rendant compte de la vitesse radiale des n{\'{e}}buleuses extra-galactiques}.
\newblock {\em Annales de la Soci{\'{e}}t{\'{e}} Scientifique de Bruxelles},
  A47:49--59, 1927.

\bibitem{mach_2013}
Ernst Mach.
\newblock {\em {The Science of Mechanics: A Critical and Historical Exposition
  of its Principles}}.
\newblock Cambridge Library Collection - Physical Sciences. Cambridge
  University Press, 2013.

\bibitem{Magueijo2019}
João Magueijo and Lee Smolin.
\newblock {A universe that does not know the time}.
\newblock {\em Universe}, 5(3):1--14, 2019.

\bibitem{Milne1933}
E.A. Milne.
\newblock {World-Structure and the Expansion of the Universe}.
\newblock {\em Zeitschrift f{\"{u}}r Astrophysik}, 6:1, 1933.

\bibitem{ORaifeartaigh2017}
Cormac O’Raifeartaigh, Michael O’Keeffe, Werner Nahm, and Simon Mitton.
\newblock {Einstein’s 1917 static model of the universe: a centennial
  review}.
\newblock {\em European Physical Journal H}, 42(3):431--474, 2017.

\bibitem{Perlmutter1999}
S.~Perlmutter, G.~Aldering, G.~Goldhaber, R.~A. Knop, P.~Nugent, P.~G. Castro,
  S.~Deustua, S.~Fabbro, A.~Goobar, D.~E. Groom, I.~M. Hook, A.~G. Kim, M.~Y.
  Kim, J.~C. Lee, N.~J. Nunes, R.~Pain, C.~R. Pennypacker, R.~Quimby,
  C.~Lidman, R.~S. Ellis, M.~Irwin, R.~G. McMahon, P.~Ruiz‐Lapuente,
  N.~Walton, B.~Schaefer, B.~J. Boyle, A.~V. Filippenko, T.~Matheson, A.~S.
  Fruchter, N.~Panagia, H.~J.~M. Newberg, W.~J. Couch, and The
  Supernova~Cosmology Project.
\newblock {Measurements of {$\Omega$} and {$\Lambda$} from 42 High‐Redshift
  Supernovae}.
\newblock {\em The Astrophysical Journal}, 517(2):565--586, 1999.

\bibitem{Planck2015}
{Planck Collaboration}.
\newblock {Planck 2015 results}.
\newblock {\em Astronomy {\&} Astrophysics}, 13, 2016.

\bibitem{Riess1998}
Adam~G. Riess, Alexei~V. Filippenko, Peter Challis, Alejandro Clocchiatti, Alan
  Diercks, Peter~M. Garnavich, Ron~L. Gilliland, Craig~J. Hogan, Saurabh Jha,
  Robert~P. Kirshner, B.~Leibundgut, M.~M. Phillips, David Reiss, Brian~P.
  Schmidt, Robert~A. Schommer, R.~Chris Smith, J.~Spyromilio, Christopher
  Stubbs, Nicholas~B. Suntzeff, and John Tonry.
\newblock {Observational Evidence from Supernovae for an Accelerating Universe
  and a Cosmological Constant}.
\newblock {\em The Astronomical Journal}, 116(3):1009--1038, 1998.

\bibitem{Robertson1929}
H.P. Robertson.
\newblock {On the Foundations of Relativistic Cosmology}.
\newblock {\em Proc N.A.S.}, 15:822--829, 1929.

\bibitem{Rovelli1991}
Carlo Rovelli.
\newblock {Time in quantum gravity : An hypothesis}.
\newblock {\em Physical Review D}, 43(2), 1991.

\bibitem{Rovelli2008}
Carlo Rovelli.
\newblock {Forget Time}, 2008.

\bibitem{Rovelli2017}
Carlo Rovelli.
\newblock {\em {The Order of Time}}.
\newblock Allen Lane, 2017.

\bibitem{Springel2006}
V~Springel, C~Frenk, and S~White.
\newblock {The large-scale structure of the Universe}.
\newblock {\em Nature}, pages 1--34, 2006.

\bibitem{Steinhardt2021}
Charles~L. Steinhardt, Christian~Kragh Jespersen, and Nora~B. Linzer.
\newblock {Finding High-redshift Galaxies with JWST}.
\newblock {\em The Astrophysical Journal}, 923(1):8, 12 2021.

\bibitem{VandenBergh1999}
Sidney van~den Bergh.
\newblock {The Early History of Dark Matter}.
\newblock {\em Publications of the Astronomical Society of the Pacific},
  111(760):657--660, 1999.

\bibitem{Walker1935a}
A.~G. Walker.
\newblock {On riemanntan spaces with spherical symmetry about a line, and the
  conditions for isotropy in genj relativity}.
\newblock {\em Quarterly Journal of Mathematics}, os-6(1):81--93, 1935.

\bibitem{Will2014}
Clifford~M. Will.
\newblock {The confrontation between general relativity and experiment}.
\newblock {\em Living Reviews in Relativity}, 17, 2014.

\end{thebibliography}

\end{document}